\documentstyle[BIB]{report}    
\renewcommand{\theequation}{\thesection.\arabic{equation}}
\newcommand{\be}{\begin{equation}}
\newcommand{\ee}{\end{equation}}
\def\lsim{\raise0.3ex\hbox{$<$\kern-0.75em\raise-1.1ex\hbox{$\sim$}}}
\def\gsim{\raise0.3ex\hbox{$>$\kern-0.75em\raise-1.1ex\hbox{$\sim$}}}

\title{Four-quark Binding Energies from SU(2) Lattice \\
Monte Carlo}
\author{ A.M. Green\thanks{E-mail address: green@phcu.helsinki.fi},\\
 Research Institute for Theoretical Physics,  University
of Helsinki, Finland, \\
 C. Michael\thanks{E-mail address: cmi@liverpool.ac.uk},\\
DAMTP, University of Liverpool, UK, \\
 and M.E. Sainio\thanks{E-mail address: sainio@phcu.helsinki.fi}\\
 Dept. of Theoretical Physics, University of Helsinki, Finland.}

\begin{document}

\large

\maketitle

\begin{abstract}

  Energies of four-quark systems have been extracted in a static quenched
SU(2) lattice Monte Carlo calculation for six different geometries, both planar
and non-planar, with $\beta=2.4$ and lattice size $16^3\times 32$.
In all cases, it is found that the binding energy is greatly enhanced when
the four quarks can be partitioned in two ways with comparable energies.
Also it is shown that the energies of the four-quark states cannot be
understood simply in terms of two-quark potentials.

\end{abstract}

\newpage

\section{ Introduction}
\setcounter{equation}{0}

In the literature there are several attempts to describe
meson-meson scattering \cite{MMS} and the nucleon-nucleon potential \cite{NN}
as the interaction between clusters of quarks -- $(q_1\bar{q}_3)(q_2\bar{q}_4)$
and $(q^3)(q^3)$ in these two cases. However, the temptation has been to
evaluate these interactions in terms of the \underline{two}-quark potential
acting between pairs of quarks.
Of course, for similar problems in other branches of physics this is well
justified. For example, in atomic physics, the use of the Coulomb
potential between
all \underline{pairs} of charged particles is a very good approximation.
Similarly, in nuclear physics many phenomena can be described quite well
in terms of
simply the nucleon-nucleon potential or an effective \underline{two}-nucleon
interaction derived from this. However, for clusters of quarks interacting
via gluons such simple rules have yet to be justified.

In an attempt to clarify the form of the effective interaction between
quarks in multi-quark systems, a series of model calculations have recently
been undertaken \cite{MGP}-\cite{GBP}. At this stage, it is probably far too
ambitious to contemplate a realistic calculation of, for example, the complete
nucleon-nucleon potential, since this would involve questions in addition
to the main one being addressed here, namely, the magnitude and form of the
effective quark interaction\footnote{\large{However, this has not deterred
studies of the subject e.g.\cite{Markum}.}}.
Such questions concern how the long range part
of the NN potential -- known to be well described in terms of meson
exchange -- can be quantitatively reproduced in terms of quarks and also how
this region in configuration space can be joined to the short range part
that is expected to be governed by explicit quark interactions \cite{Vinhmau}.
In order to avoid these problems and to focus attention on the quark dominated
domain, in this paper a model calculation is considered that has these
features.
More explicitly, the energies of four quarks in specific geometries are
calculated in the static quenched approximation using the SU(2) gauge group.
The "static" approximation implies that the quarks are infinitely heavy
i.e. they are treated as fixed colour sources, and their kinetic
energy neglected -- and "quenched" implies  the creation and annihilation of
$q\bar{q}$ pairs is also neglected -- details being given in
refs.\cite{GMP1}-\cite{GBP},\cite{CM1}-\cite{Hunt}.
As seen from ref.\cite{chrism}, the restriction
to SU(2) is not expected to lead to qualitative differences compared with
SU(3), whereas the computer effort is roughly an order of magnitude smaller.

  Since the aim of the present work is to evaluate the energies of the
four-quark states \underline{exactly} and not in some perturbative manner,
a Monte Carlo simulation is employed. However, as will be seen later, the main
quantity of interest -- the \underline{binding} energy of the four-quark
state -- is essentially the \underline{difference} between two energies that
are comparable. This results in extracting quantities that are small due to
substantial cancellations. In fact, many geometries result in binding energies
that are essentially zero within the accuracy of the simulation. Therefore,
an attempt is made to discuss mainly those cases where the cancellation
is less dramatic. Since these situations are not commonly met in Monte Carlo
lattice simulations, special care is needed to ensure that the resulting
binding energies do not contain any significant lattice artefacts.
In section 2, a brief description of the Monte Carlo simulation is given --
with
emphasis on those aspects relevant to the present calculation.

A by-product of these four-quark calculations are the corresponding
colour-singlet two-quark
potentials $(V_2)$. These are a necessary ingredient for any model that
attempts
to explain 4-quark energies in terms of combinations of various $V_2$'s and so
they will be discussed in some detail in section 3. Of course, when
ignoring spin, the most general form of $V_2$ on a lattice depends, at most,
on three coordinates $x,y,z$ i.e. $V_2(x=x_2-x_1,y=y_2-y_1,z=z_2-z_1)$,
where the ($x_i,y_i,z_i$) are the lattice coordinates of quark $i$. If the
distance between the two quarks is a sufficiently large number of lattice
steps, then approximate rotational invariance is expected e.g.
$V_2(5,0,0)\approx V_2(3,4,0)$ and $V_2(3,0,0)\approx V_2(2,2,1)$. This will
be illustrated.

With four quarks it is not possible to do an exhaustive study that covers
all combinations involving interquark distances upto, say, 5--10 lattice
spacings. Therefore, a selection is made of different geometries each of which
has a characteristic feature. For simplicity, all geometries have the
constraint that the lowest energy configuration is composed of two "mesons"
of equal length $d$.
In this way, it is hoped that the energies of 4-quarks
in other geometries can be reliably predicted. The specific geometries
considered are those shown in fig.1.
\begin{itemize}
\begin{enumerate}
\item The rectangles(R) of fig. 1a) include the special case of squares. The
latter are found to have the largest binding energies ($\approx 100$ MeV) of
all the geometries treated here -- a feature attributed to the fact that the
basic configurations $A$ and $B$
are degenerate in energy. This conclusion
is supported in other geometries by the observation of increased binding in
those situations where the lowest energy basic configurations are almost
degenerate in energy. To study this
point further, a series of independent runs for only squares is carried out.
This includes larger squares (upto $7\times 7$) than the original
rectangle series. This will be refered to as the Large Square(LS) geometry.
\item Fig. 1b) shows the Tilted Rectangle (TR) geometry. This is of interest
for two reasons:
\begin{itemize}
\item a) The areas of the rectangles subtended by the 4-quarks are
intermediate to those in fig. 1a) -- permitting a more reliable parametrization
of the binding energy for 4-quarks in a general rectangular geometry.
  Only those configurations are treated, in which $A$ and $B$ are as
degenerate as possible in energy. This leads to binding energies that are
significantly different from zero.

\item b) The special case $d=5,x=4,y=3$ leads to a square with sides of
length 5 lattice units. The comparison with the corresponding $5\times 5$
square in fig. 1a) should give a measure of the goodness of rotational
invariance in the 4-quark case for this particular configuration size.
\end{itemize}
\item The Linear(L) case in fig. 1c) is of interest for two reasons:
\begin{itemize}
\item a) Unlike the two above rectangular geometries, where the "natural"
selection of basis states was $A$ and $B$, here no such choice is obvious.
Therefore, all three states $A,B$ and $C$ are considered .
The consequences of this are discussed further in Appendix A.

\item b) As shown in ref.\cite{GMP2} a comparison can be made with QCD in two
dimensions (1+1).

\end{itemize}
\item Fig. 1d) shows the so-called Quadrilateral(Q) geometry, which can be
thought of as an intermediate situation between the Rectangle and Linear
cases. Again the "natural" choice of basis is not obvious, so that all three
states $A,B$ and $C$ are taken into account.
As discussed in Appendix A, the third state adds very little to
the energy determinations.
\item  Fig. 1e) depicts a Non-Planar(NP) geometry. This is related to
the Rectangle(R) and Quadrilateral(Q) geometries. The Q-geometry can be
viewed as a transformation of the R-geometry, in which one of the two-quark
clusters $(q_2q_4)$ in state $A$ is rotated through $\pi/2$
about one of the quarks in the
plane of the rectangle. In this way the NP-geometry is a similar
transformation, in which the same quark cluster is rotated by $\pi/2$
perpendicular to the original R-plane.
\end{enumerate}
\end{itemize}
The energies corresponding to these six geometries are given in section 4.
Finally, in section 5, some conclusions are drawn by considering a simplified
model of four quarks interacting via purely two-quark potentials.

\section{ Outline of the Monte Carlo Simulation}
\setcounter{equation}{0}

 \vskip 0.5 cm

Since the basic techniques for performing the type of Monte Carlo
simulation used in this work are well documented \cite{CM1}-\cite{chrism},
only a brief description will be given here.

All calculations are carried out with a $16^3\times 32$ lattice and
$\beta=2.4$ -- a value appropriate for SU(2) and corresponding to a lattice
spacing of $a\approx 0.12$ fm. In a previous paper \cite{GMP2}, the effect
of scaling
and lattice finite size were checked on a $24^3\times 32$ lattice with
$\beta=2.5$. The lattices are generated by a standard Wilson action
employing a heat bath method for equilibration. Further updates are
combinations    of three over-relaxation sweeps and one
heat bath sweep. The measurement of the appropriate correlation functions
takes place always after such a cycle of four sweeps.
In order to make error estimates, successive
measurements are collected into a series of blocks.

The measurements made on these lattices are of two types -- the extraction
of two-quark potentials ($V_2$) and the extraction of four-quark potentials.

\subsection{Two-quark potentials}

To extract two-quark potentials, the basic lattice links produced after
each series of four updates are "fuzzed".
  Essentially this models a glue flux-tube between static colour
sources. As discussed in refs.\cite{CM1}-\cite{CM2}, this is an iterative
procedure, which replaces a given spatial link by gauge invariant
combinations of the nearest neighbouring spatial links. In line with
refs.\cite{GMP}-\cite{GMP2}, the factor that weights the original link
with respect to the combination of neighbouring links is taken to be $c=4$.
This procedure  can be repeated many times -- the number of fuzzing levels.
To measure
a two-quark potential $V_2(r)$ along a given spatial axis, the two quarks are
connected by a \underline{straight line} path $P_i$ through a series of
links -- all fuzzed to the same level. Different paths can then be constructed
using different fuzzing levels  -- 12, 16 and 20 in the present work. As will
be shown below, this results in a $3\times 3$ eigenvalue problem with the basis
$P_1=P(12 \ {\rm fuzzes}) \ , \ P_2=P(16 \ {\rm fuzzes})$ and
$P_3=P(20 \ {\rm fuzzes})$.

  Given these three paths $P_i$,
then the correlation between paths $P_i$ and $P_j$ -- parallel
transported by $T$ steps in the time direction -- is
\be
\label{WP}
W^T_{ij} = <P_i| \hat{T}^T | P_j>,
\ee
where $\hat{T} = \exp {(-a\hat{H})}$ is the transfer matrix for one time
step with the Hamiltonian $\hat{H}$. A trial wave function
$\psi = \sum_{i} a_i |P_i>$ leads to the $3\times 3$ eigenvalue equation
\be
\label{WABC}
W^T_{ij} a^T_j = \lambda^{(T)} W^{T-1}_{ij} a^T_j,
\ee
where $\lambda^{(T)} \rightarrow \exp{(-aV_{2})}$ as $T \rightarrow \infty $
and which, for a single state, reduces to the form
\be
V_2(r)=-\frac{1}{a} \lim_{t \rightarrow \infty}
\ln \left( \frac{W(r,T)}{W(r,T-1)} \right).
\ee
Besides giving the possibility to extract also higher eigenvalues
in addition to the ground state, the method yields accurate information
about $V_2$ already for small values of $T$ ($T \leq 5$) \cite{Hunt}.

To measure two-quark potentials not along a given spatial axis i.e. $V_2(x,y)$
and $V_2(x,y,z)$, again three paths are constructed, each of which has 12, 16
and 20 fuzzings. However, each path is itself a combination of paths -- all
with the same fuzzing level $i$. For $V_2(x,y)$, the two paths $P_i(a,b)$
shown in fig. 2 are combined with equal weight i.e.
\be
\label{Pxy}
P_i(x,y)=\frac{1}{2}[P_i(a)+P_i(b)].
\ee
For the fully 3-dimensional case, $V_2(x,y,z)$, combinations around the
sides of a cuboid are made. As seen in fig. 1, for configuration $B$ of the
NP-geometry, the path of interest goes from point 1 at $(0,0,z_0)$ to point 4
at $(x_0,y_0,0)$ -- with points 5 and 6 at ($x_0,0,z_0$) and ($0,y_0,z_0$) not
 shown. This is taken to be the combination

\vskip 0.5 cm

\[P_i(1\rightarrow 4, xyz)=\]
\be
\label{PP}
\frac{1}{3}[P_i(1\rightarrow 5, x)P_i(5\rightarrow 4, yz)+
P_i(1\rightarrow 6, y)P_i(6\rightarrow 4, xz) +
P_i(1\rightarrow 3, z)P_i(3\rightarrow 4, xy)],
\ee
where, for example, $P_i(1\rightarrow 5, x)$ is parallel to the $x$-axis
and $P_i(5\rightarrow 4, yz)$ is itself a combination of two
paths in the $yz$-plane -- as in eq.(\ref{Pxy}).   It should be pointed out
 that,
in eq.(\ref{PP}), the product of two paths $P_i$ is to be interpreted as
a product of two SU(2) matrices.

The reason for this rather detailed discussion of how paths are constructed
in the off-axis cases will be apparent later, when the results from the Linear
   and
Quadrilateral geometries are given. There it will be seen that the particular
method used in this work has interesting consequences.

\subsection{Four-quark potentials}

When measuring the 4-quark potentials,
 the paths $P_i$ in eq.(\ref{WP}) now refer to the 4-quark states
$A$, $B$ and -- possibly -- $C$ of fig. 1. In some cases, such as
the Linear geometry, these three states can be linearly dependent -
this is discussed in Appendix A. Again, those paths involving pairs of
 quarks
that are not along a spatial axis are treated as in eq.(\ref{Pxy}) or
(\ref{PP}). In principle, each of the 4-quark states could also have a
"fuzzing" index. However, here only the maximum fuzzing level of 20 is used.
Including several "fuzzing" levels would add little in
accuracy and is computationally demanding.
A $2\times 2$ or $3\times 3$
correlation matrix in eq.(\ref{WABC})  is found to be  adequate.

In order to evaluate the statistical errors on the various 2- and 4-quark
energies, it is convenient to collect $M$ successive measurements into
a single block i.e. for a given geometry, $N$ blocks containing $N\times M$
measurements are
generated by $4\times N\times M$ sweeps. In ref.\cite{GMP2}, by evaluating
the autocorrelation function, it is shown that these blocks are essentially
independent
of each other.
  Given these $N$ blocks, the values of the energies -- in units of the
inverse lattice spacing --  can be extracted by the simple averaging
\be
\bar{E}^T=-\ln\left[\sum^N_{m=1} W_m(T+1)/\sum^N_{m=1} W_m(T)\right],
\ee
where $W_m(T)$ is the average of the
$M$ measurements of the Wilson loop in block $m$ for a Euclidean
time difference $T$.
However, the errors on these averages are estimated using
the Bootstrap method \cite{Boot} , in which the basic set of $N$ blocks is
replaced
by $N_B$ sets of $N$ blocks. These $N_B$ sets are created from the basic set
by a random selection with replacement.
For each of these $N_B$ sets of $N$ blocks an average energy $\bar{E}_j$
($j=1,...,N_B$) is calculated.
The root mean square deviation $\sigma_B$ of these $\bar{E}_j$ (Euclidean
time index is suppressed for these quantities)
is then taken to be an estimate of the error in the energy $\bar{E}^T$ from
the original $N$ blocks. In other words, the energies from the
\underline{original} blocks are considered to be the best estimates, but the
errors on these estimates are given by the \underline{Bootstrap} value
$\sigma_B$.
 In this case $N_B=499$ is
chosen. In table 1 are given the number of blocks$(N)$ and the number of
measurements$(M$) within each block for each of the geometries being discussed.
In addition, for each geometry, the number of four-quark    possibilities
considered (e.g. in fig.1 $r,d=1,..,4$ with $r\geq d$)   and
also the approximate CPU time on the Helsinki CRAY X-MP are shown.
\begin{table}[t]
{Table 1: For each of the six geometries is shown:
a) The number of blocks $N$, \newline b) The number of measurements per
block $M$,
c) The total number of measurements ($N_T=N\times M$),
d) The number of  $(d,r)$-values considered $N_C$,
e) Approximate total CPU time (in hrs) on the Helsinki CRAY X-MP.}
\begin{center}
\vskip 0.5 cm
\begin{tabular}{|c|c|c|c|c|c|} \hline
Geometry&$N$&$M$&$N_T$&$N_C$&CPU(hrs) \\ \hline
Rectangles(R)&10&80&800&10&$\approx$100 \\
Large Squares(LS)&10&160&1600&7&$\approx$150\\
Tilted Rectangles(TR)&15&96&1440&10&$\approx$100\\
Linear(L)&10&160&1600&9&$\approx$200\\
Quadrilaterals(Q)&12&60&720&9&$\approx$100\\
Non-Planar(NP)&10&96&960&16&$\approx$150\\ \hline
\end{tabular}
\end{center}
\end{table}
As will be discussed later in connection with table 4, the set of $N$
blocks from a single geometry can -- for some purposes such as evaluating
two-quark potentials -- be considered as a
single "super-block". In this way the \underline{same} two-quark potential
is calculated in the context of different geometries,
so that a combined analysis of these
super-blocks will help towards an understanding  of the question of
systematic errors.

In the previous paragraph, the "energies" are referred to in several places.
However, as pointed out after eq.(\ref{WABC}), the term "energy" should be
reserved for
the asymptotic limit as the number of Euclidean time steps $T\rightarrow
\infty$.
In order to extract this limit, for each geometry there are measurements
$\bar{E}^{T}=-\ln(\lambda^{T})$ for $T=1,2,3,4,5$,
where -- from now on -- all energies
will be given in units of $1/a$.  In the case of the 2-    and
4-quark potentials ($V_2$ and $V_4$), the $\bar{E}^{T}$ should
monotonically decrease  to a constant -- the desired energy $E$ -- for
sufficiently large T.  Unfortunately, by $T=5$, the errors $\sigma_B$
generated
by the Bootstrap method can become as large as the $\bar{E}^{T}$'s themselves.
Therefore,
it is necessary to develop strategies for extracting the desired asymptotic
value $E$. In the present work, two such strategies are used.   The first
strategy attempts to identify $E$ as a "plateau" in the $\bar{E}^T$ by
comparing
successive values of the $\bar{E}^T$'s. On the other hand, the
second strategy resorts to an optimized "overall fit" to the $\bar{E}^T$'s.
They are best illustrated by table 2.
\begin{table}[t]
{Table 2: Strategies for extracting energies -- see text for explanation.}
\begin{center}
\vskip 0.5 cm
\begin{tabular}{|c|c|c|c|c|c|} \hline
T&A&B&C&D&F \\ \hline
1&$A_1$&$B_1$&$C_1$&& \\
2&$A_2$&$B_2$&$C_2$&$D_2$&$F_2$ \\
:&:&:&:&:&: \\
5&$A_5$&$B_5$&$C_5$&$D_5$&$F_5$ \\   \hline
\end{tabular}
\end{center}
\end{table}
\noindent Here column A shows, for a given Euclidean time step $T$,
the values $<\bar{E}_j>$ averaged over the the
$j=1,...,N_B$ Bootstrap blocks and column C the corresponding errors
$\sigma_B$. On the other hand, column B shows the average $\bar{E}^T$ from the
\underline{original} $N$ blocks. Of course, as a sign of consistency, the
entries in columns A and B should be almost equal.   This is
 indeed the situation
in the present problem. If there had been a difference, then it could have
been corrected for as discussed in ref.\cite{cdata}.
Column D is simply the
difference between successive values of $B_i$ i.e.
\mbox{$D_i=B_{i-1}-B_i$}, and
column F is the corresponding standard deviation
of these differences obtained from the Bootstrap analysis in column A.
 The problem is how to extract from
column B the asymptotic value. In the first strategy, the value of $T=T_0$ for
which $D_{T_0}\pm F_{T_0}$ is consistent with zero is recorded, whenever
possible. The best estimate of the asymptotic energy is then taken to be
$B_{T_0-1}\pm C_{T_0-1}$. In the second strategy, the values of
$B_T\pm C_T$ are fitted for $T=2,..,5$ -- using the Minuit programme \cite
{Roos} -- with
the function $\ln\left[G(T)/G(T+1)\right]$ where
\be
\label{GT}
G(T)= \exp(-E_M^{(0)}T)+L_M\exp(-E_M^{(1)}T).
\ee
The rational behind this form of $G(T)$ is seen in eqs.(\ref{WABC}) and (2.3).
There, for small
values of $T$, the desired ground state energy $E_0$ is
contaminated by higher states. This is also the reason for not including
$B_1 \pm C_1$ in the fitting procedure.
The value of $E^{(0)}_M$ is then expected to be
a good estimate of $E_0$   with $E^{(0)}_M$ approaching $E_0$
 monotonically
from above. Thus $L_M$ should be positive.
 However, it is probably not possible to interpret
$E^{(1)}_M$ as an estimate of the energy of some particular excited state,
since -- for small $T$ -- it must be simulating the effect of many such excited
states. Also, $E^{(1)}_M$ should be constrained not to be too close to
$E^{(0)}_M$.
The interesting feature relating these two strategies is that
in general they give
essentially the \underline{same} value with similar errors  for the best
estimate of the ground state energy i.e.
\be
E^{(0)}_M\approx B_{T_0-1} \pm C_{T_0-1}
\ee
and gives confidence in the first strategy  that, indeed, the choice
$B_{T_0-1}\pm C_{T_0-1}$ is more realistic than the possible -- but more
pessimistic -- choice $B_{T_0}\pm C_{T_0}$.
This is important, since the
second strategy, based on eq.(\ref{GT}), is strictly speaking more
appropriate to the \underline{total} energies $V_2$ and $V_4$ of the
2-and 4-quark states, but less so for the \underline{binding}
energy of the 4-quark state $(BE)$ because the latter is the
\underline{difference}
\be
\label{diff}
BE=\left[V_4-V_2(q_1q_3)-V_2(q_2q_4)\right]=\left[V_4-2V_2(d)\right]
\ee
in the notation of fig.1.
 The second stage in this equation arises,
since in this work only geometries are chosen, where the basis state of lowest
energy is composed of "mesons" of equal length $d$ i.e.
$V_2(q_1q_3)=V_2(q_2q_4)=V_2(d)$.
The reason for the form of eq.(\ref{diff}) will
become more apparent in the next section. Since the difference $BE$ can be
small,
more care is now needed when making the $T$-extrapolation.

In practice, strategy 2 is applied in two ways:
\begin{itemize}
\begin{enumerate}
\item After extracting from $V_2$ and $V_4$ -- both of which are taken from
the \underline{same} Monte Carlo configurations to reduce the error on the
difference --
 the values of $E_M^{(0)}$(2-body) and
$E_M^{(0)}$(4-body) \underline{separately}, eq.(\ref{diff}) then gives
directly an
estimate of $BE$ as a difference of these $E_M^{(0)}$ values. This
\underline{always} gives quantities in good agreement with strategy 1 for
$BE$ -- see $E_0^*$ in table 8. Furthermore, eq.(\ref{GT}) can also be used
to extract
an estimate of the energy of the \underline{first excited} state of the
4-quark system i.e. $V_4$(excited)$\rightarrow E_M^{(1)}$(4-body).
However, as mentioned before, the use of eq.(\ref{GT}) is less justified
theoretically.
 Evenso, the resultant predictions $E_1^*$ in table 8 are still found to be
in very good agreement with strategy 1. The
disadvantage of this approach is that it is not possible to get a reasonable
estimate of the error on $BE$, since this would require combining the separate
errors on the various $E_M^{(0)}$'s. An attempt to avoid this problem
is as follows.
\item Use eq.(\ref{GT}) to extrapolate directly the expression for $BE$ in
eq.(\ref{diff}) or the corresponding difference involving the first excited
state. In both of these cases, the form of eq.(\ref{GT}) is even less
justified theoretically.  Surprisingly, in practice, it is found that this
second method can also produce estimates $(E^{(0)}_M)$ for $BE$ that are
still in agreement   with $B_{T_0-1} \pm C_{T_0-1} $. In fact, the method
often works even better for the excited state energy $E_1$.
\end{enumerate}
\end{itemize}

\noindent The conclusion is that, in no cases do the two strategies
suggest significantly different estimates for either $E_0$ or $E_1$,
which gives confidence in the reliability of the values of these two
energies.

As an example of these ideas, in table 3 the actual numbers for the total
4-quark energy ($V_4)$ of a $5\times 5$ square are given.
\begin{table}[t]
{Table 3: Example of table 2 for the total energy $(V_4)$ of a
 $5\times 5$ square.
The underlined numbers in columns $B$ and $C$ are the best estimate of the
ground state energy and its error -- see text for a detailed explanation.}
\begin{center}
\vskip 0.5 cm
\begin{tabular}{|c|c|c|c|c|c|} \hline
T&A&B&C&D&F \\ \hline
 1&  1.76835 & 1.76838 & 0.00107&&        \\
 2&  1.68648 & 1.68653& 0.00160 &  0.08185&  0.00078 \\
 3&  1.67462 &  \underline{1.67476}&   \underline{0.00225}&  0.01177&
 0.00123 \\
 4&  1.66809 & 1.66854&  0.00877& \underline{0.00622}&  \underline{0.00893} \\
 5&  1.64386 & 1.64369&  0.03541&  0.02485&  0.03842\\ \hline
\end{tabular}
\end{center}
\end{table}
\noindent Here it is seen that the first strategy gives convergence at $T=3$
since
$D_4\pm F_4$ is consistent with zero. In this case the second strategy
yields $E^{(0)}_M=1.670(3)$ -- showing that
$E^{(0)}_M$ agrees well with the 1.675(2) from table 3. However, it should
be added that the situation is often less clear cut when applying these
strategies to extract the 4-quark binding energies $BE$. Evenso, as will
be later seen in
table 8, the agreement between the two strategies is in general
still maintained.

\section{ Two-Quark Potentials}
\setcounter{equation}{0}

In any attempt to understand four-quark binding energies  the values of the
corresponding
two-quark potentials $(V_2)$ could well play a role. This will be illustrated
in section 5, where a brief outline of the method advocated in
ref.\cite{MGP}-\cite{GBP} will be given. Therefore, in this section a
discussion is made
 of the values of $V_2$ that arise when calculating the 4-quark energies for
the various geometries of fig.1.
  As shown, for example in refs.\cite{CM1}-\cite{CM2}, the present
techniques result in very accurate measurements of $V_2$ -- with the
errors estimated by the above Bootstrap method being, in most cases, at the
$0.1\%$ level. In spite of this accuracy (or perhaps because of it) some values
of $V_2$ differ from what, at first sight, is expected. Such cases will
be discussed below.

\vskip 0.5 cm

\subsection{Comparison of the same $V_2$'s between different geometries.}

\vskip 0.25 cm

In this work some of the $V_2$'s are estimated
in several independent runs  -- for example --
$V_2(r)$ with $r=1,2 \ \rm{and} \ 3$ is calculated for all geometries and
$V_2(x=4,y=3)$ is calculated in the Tilted Rectangle and Quadrilateral
geometries. Therefore, as discussed after table 1, the results from a given
geometry can be considered as a single super-block of data.
This then permits an evaluation of systematic errors to be made by
comparing these different super-blocks, since the same method is used to
extract the separate estimates of $V_2$.
A typical example of this is $V_2(r=5)$, which is calculated
in all geometries -- except the Non-Planar case -- and, for the two strategies,
takes on the values in table 4.
\begin{table}[t]
{Table 4: The value of $V_2(r=5)$ from the series of runs for
the various geometries -- using
the two strategies of section 2.
$V_2({\rm Average})=\bar{V}_2=$
\mbox{$\sum_{j=1}^5 N^j_TV^j_2(r=5)/\sum_{j=1}^5 N^j_T$}, where $j$
   denotes the 5 separate geometries and the $N^j_T$'s are given in table 1.
The quoted error contains a factor that is $\approx 1/\sqrt{5}$, since it is
the error on the average that is needed -- see ref.\cite{LL} for the
actual expression used.}
\begin{center}
\vskip 0.5 cm
\begin{tabular}{|c|c|c|c|c|c||c|} \hline
Strategy&R&LS&TR&L&Q&$\bar{V}_2$\\ \hline
1&0.8583(9)&0.8576(16)&0.8563(17)&0.8562(16)&0.8545(14)&0.8567(6) \\
2&0.8583(7)&0.8576(22)&0.8562(19)&0.8556(23)&0.8540(13)&0.8564(7)\\ \hline
\end{tabular}
\end{center}
\end{table}
\noindent Here it is seen that the spread of values from the Rectangle(R)
case to
the Quadrilateral(Q) case is 0.0038 and 0.0043 for the two strategies
respectively  -- several times larger than the
standard deviations estimated by these strategies and which are shown
in parentheses. This is a systematic trend, in the sense that, in the present
work, the Quadrilateral results also
for other values of $r$ are always smaller than the estimates from the
other geometries. To a large extent these fluctuations between different
geometries are avoided, when calculating the 4-quark binding energies $BE$,
since the difference in eq.(\ref{diff}) is calculated directly for each
geometry.
This appears to be quite well justified -- as will be seen
later, in subsection 3c), when a comparison is made between a Tilted Rectangle
of area $5\times 5$ and a Large Square of the same dimensions.

The reasons for the above discrepancies in table 4 are not easy to pin-point.
However, three possibilities are as follows:

i) For a given geometry, standard
autocorrelation tests, as in ref.\cite{GMP2}, suggest that
a more realistic standard deviation, $\sigma_B(Auto)$, is given by
$\sigma_B\sqrt{f_c}$, where $\sigma_B$ is the Bootstrap value discussed in
table 2. However, it is found that $f_c$ is never larger than about 1.5 -- a
correction that is too small to account for the above fluctuation.
  This means that \underline{within} each super-block there is no
statistically significant evidence of autocorrelation.

ii) When the distributions of the averages from each of
the $N_B$ Bootstrap blocks are plotted, they all seem to be
quite well described by a gaussian of the same RMS distribution. Some of the
cases may give the hint of a slight skewness or non-gaussian like tails, but
certainly not sufficient to rule out the use of standard statistical ideas
\underline{within} each super-block.

iii) Since the problem does not seem to lie within the super-blocks, it
indicates that there perhaps exists some long-range correlation
\underline{between} the super-blocks.

However, to get a correct perspective of these discrepancies, it should be
observed
that the weighted averages $\bar{V}_2$ of each row in table 4 differ from the
individual entries  by at most 3-4$\sigma$ -- a not improbable  situation.
It should also be pointed out that $V_2(Q)$  --  the entry furthest
from $\bar{V}_2$ -- contains less measurements than the other geometries -- see
$N_T$ in table 1.
Therefore, probably the observed fluctuation is a \underline{combination} of
these various effects and, as such, can be considered quite acceptable.

\vskip 1.0 cm

\subsection{A global average form of $V_2$.}

\vskip 0.5 cm

For completeness, in this paper a \underline{global} average $V_2$ is given
in table 5. This is constructed in the following way using only
the results from strategy 1 -- the difference with strategy 2 being
minute.
In those cases, where there are multiple estimates of a given
$V_2$, then their weighted average and standard deviation are first calculated
 from these different estimates e.g. from the first row of table 4,
$\bar{V}_2(r=5)$ is taken    to be 0.8567(6).
As was found to be sufficient in ref.\cite{GMP2}, only
the weighted averages of $V_2(r=2,3,4,5,6)$ over the five super-blocks were
utilized.
These  are now fitted with the Minuit routine\cite{Roos} to give
the parameters e=0.250(5), $a^2b_s$=0.0703(6)  and $av_0$=0.555(4)
in  the lattized form of
\be
\label{Coulomb}
V_2(r)=-(e/r)_L+b_sr+v_0.
\ee
\begin{table}[t]
{Table 5: A global fit to the two-quark potentials $V_2$.
The numbers underlined are those used in the fit.}
\begin{center}
\vskip 0.5 cm

\begin{tabular}{|c|c|c||c|c|c|} \hline
 $(\frac{\displaystyle x}{\displaystyle a},\frac{\displaystyle y}
{\displaystyle a})$&$V_2(MC)$&$V_2(Fit)$&
 $(\frac{\displaystyle x}{\displaystyle a},\frac{\displaystyle y}
{\displaystyle a},\frac{\displaystyle z}{\displaystyle a})$&
$V_2(MC)$&$V_2(Fit)$ \\ \hline \hline
(1,0)&0.3732(1)&0.3563&(1,1,0)&0.4885(1)&0.4823\\
(2,0)&\underline{0.5621(2)}&0.5621&(1,1,1)&0.5467(5)&0.5411\\
(3,0)&\underline{0.6803(4)}&0.6804&(1,2,0)&0.6022(3)&0.6005\\
(4,0)&\underline{0.7737(6)}&0.7734&(1,2,1)&0.6314(7)&0.6279\\
(5,0)&\underline{0.8567(6)}&0.8566&(1,3,0)&0.6989(7)&0.6982\\
(6,0)&\underline{0.9348(13)}&0.9353&(1,3,1)&0.717(1)&0.7137\\
(7,0)&1.013(2)&1.011&(1,4,0)&0.7844(9)&0.7844\\
(8,0)&1.087(1)&1.086&(1,4,1)&0.797(2)&0.7949 \\ \hline \hline
\end{tabular}
\end{center}
\end{table}
\noindent This table shows only a few of the 2-quark potentials produced
in this work.
Mainly those involving few lattice spacings are given, since these are not
well reproduced by the expression in eq.(\ref{Coulomb}). Additional
values are given in table 6.

\vskip 1.0 cm

\subsection{Comparison between $V_2$'s that have the same interquark
distances.}

\vskip 0.5 cm

There are several cases where $V_2(r,0,0), \ V_2(x,y,0)$ and $V_2(x,y,z)$
can be compared, since either $r=\sqrt{x^2+y^2}$ or $\sqrt{x^2+y^2+z^2}$ etc.
These are shown in table 6.
\begin{table}[t]
{Table 6: Comparison of $V_2$'s  having the same interquark distance.
The numbers in brackets are from the fit given by eq.(\ref{Coulomb}).}

\begin{center}
\vskip 0.5 cm

\begin{tabular}{|c|c|c|c|c|} \hline
Entry&Geometry&$V_2(r,0,0)$&$V_2(x,y,0)$&$V_2(x,y,z)$ \\ \hline \hline
&&$V_2(5,0,0)$&$V_2(4,3,0)$& \\
&&[0.8566]&[0.8573]&\\ \hline
1&TR&0.8563(17)&0.8590(15)&\\
2&Q&0.8545(14)&0.8545(19)& \\ \hline \hline
&&$V_2(3,0,0)$&&$V_2(2,1,2)$ \\
&&[0.6804]&&[0.6848] \\ \hline
3&NP&0.6808(8)&&0.689(1) \\ \hline \hline
&&&$V_2(3,3,0)$&$V_2(1,4,1)$ \\
&&&[0.7954]&[0.7949] \\ \hline
4&NP&&0.799(2)&0.797(2) \\
5&Q&&0.796(1)& \\ \hline \hline
&&&$V_2(4,1,0)$&$V_2(2,3,2)$ \\
&&&[0.7844]&[0.7856] \\ \hline
6&TR&&0.785(1)&\\
7&NP&&&0.790(2) \\ \hline \hline
\end{tabular}
\end{center}
\end{table}
Several points can be made about these results:

i) Entry 1. The two numbers are within the statistical errors of each other.
However, their difference of 0.0027 -- if taken at its face value -- is
 partially
explained by their expected theoretical difference from eq.(\ref{Coulomb}).

ii) Entry 3. Both numbers are now not within the statistical errors of each
 other.
However, their difference of 0.008 now contains a sizeable contribution (0.004)
from the expected theoretical difference.

iii) Entries 4 and 5 are all within the statistical errors of each other -- the
small difference in entry 4 again being expected theoretically.

It should be added that, if the parameters in eq.(\ref{Coulomb}) were
fitted for each geometry separately -- and not to the overall average
$\bar{V}_2$ -- , then the lattice results would be closer to the
numbers  shown in the square brackets. However, here the interest is
not in the overall magnitude but in the differences.

\vskip 2.0 cm

\section{ Four-Quark Binding Energies}
\setcounter{equation}{0}

In this section the energies of the four-quark ground state $E_0$ and
first excited state $E_1$ are collected -- referred to as $BE$ earlier.
 These are defined as
\be
\label{Ei}
E_i=V_i(4q)-E_0(A),
\ee
where $V_i(4q)$ is the total energy of the four-quark state and $E_0(A)$
 is the unperturbed energy of the states $A$ in fig. 1. For example,
 in the $r\times d$ rectangle case with $r\geq d$, $E_0(A)$ is simply
$2V_2(d)$. Therefore, $E_0$ is best interpreted as the ground state binding
energy of the 4-quark state.   Eq.(\ref{Ei}) is essentially the same as
eq.(\ref{diff}) -- the notation change being purely for convenience, since
it is no longer necessary to specify the quark indices.

\vskip 1.0 cm

a) Rectangles(R) and Large Squares(LS).

\vskip 0.5 cm

Since the  Rectangle and Large Square geometries are very similar, they are
combined in table 7. Only the basis states $A$ and $B$ in fig. 1a) are
included, but -- as will be shown later in the case of Quadrilaterals --
the more
complete space of $A,B$ and $C$ would give very similar results.
The reason for carrying out two separate series R and LS was the observation in
series R that squares had much more binding than non-squares. Therefore, the
second series (LS) was carried out to study this point further.
\begin{table}[t]
{Table 7: The values of $E_0$ and $E_1$ for the R and LS geometries.
$E'_0$ and $E'_1$ give the predictions of a simple model discussed in the
conclusion.}
\begin{center}
\vskip 0.5 cm
\begin{tabular}{|c|c|c|c|c||c|c|} \hline
 $(\frac{\displaystyle d}{\displaystyle a},\frac{\displaystyle r}
{\displaystyle a})$& $E_0(R)$& $E_0(LS)$& $E_1(R)$& $E_1(LS)$&$E'_0$&$E'_1$  \\
   \hline
(1,1)&--0.0696(2)&--0.0694(2)&0.183(1)&0.181(1) &--0.0769&0.231 \\
(1,2)&--0.0026(2)&--&0.506(1)&-- &--0.0035&0.559 \\
(1,3)&--0.0002(3)&--&0.766(1)&--  &--0.0005&0.844 \\
(1,4)&0.0003(3)&--&0.974(3)&--&--0.0002&1.083 \\
(2,2)&--0.0585(4)&--0.0581(3)&0.143(1)&0.142(1)&--0.0698&0.209 \\
(2,3)&--0.006(1)&--&0.324(1)&--&--0.0117&0.407 \\
(2,4)&--0.0004(4)&--&0.512(1)&--&--0.0036&0.622 \\
(3,3)&--0.054(1)&--0.053(1)&0.114(1)&0.117(1)&--0.0767&0.230 \\
(3,4)&--0.006(1)&--&0.251(1)& --&--0.0245&0.385 \\
(4,4)&--0.051(1)&--0.055(3)&0.097(2)&0.097(2)&--0.0908&0.272 \\
(5,5)&--&--0.045(1)&--&0.074(4)&--0.1073&0.322 \\
(6,6)&--&--0.036(3)&--&0.059(5)&--0.1245&0.374 \\
(7,7)&--&--0.019(3)&--& 0.046(6)&--0.1430&0.429 \\
\hline
\end{tabular}
\end{center}
\end{table}
The main points of interest in table 7 are:
\begin{itemize}
\begin{enumerate}
\item The square configurations have binding energies that are an order
of magnitude larger than those of the nearest neighbouring rectangles
e.g. $E_0(3,3)=-0.054(1)$ compared with $E_0(2,3)=-0.006(1)=E_0(3,4)$.
\item For rectangles even further removed from being squares, the binding
energies are at least another order of magnitude smaller e.g.
$E_0(2,3)=-0.006(1)$ compared with $E_0(2,4)=-0.0004(4)$.
\end{enumerate}
\end{itemize}
On this evidence, it would seem to be a reasonable approximation to assume
that there is a significant interaction \underline{only} in the square
case i.e. where
states $A$ and $B$ are degenerate in energy. This can possibly be extended
to more general geometries -- a point to be considered later in this section.

\vspace{1.0cm}

b) The Tilted Rectangle(TR) geometry.

\vspace{0.5cm}

In this case, only those configurations are considered in which states $A$
and $B$ in fig. 1b) are as near as possible degenerate in energy. This is an
attempt to study how the binding energy varies with this approximate
degeneracy. For convenience, $E_0$ is now given in two forms
\be
\label{EA}
E_0^A(4q)=V_0(4q)-E_0(A)
\ee
as in eq.(\ref{Ei}), and also as
\be
\label{EB}
E_0^B(4q)=V_0(4q)-E_0(B)
\ee
in those cases, where the unperturbed energy of state $B$ is lower than that
of state $A$. It should be added that these differences are kept separate
in each of the $N$ blocks of measurements. This enables the Bootstrap analysis
to give errors for both $E_0^A(4q)$ and $E_0^B(4q)$. If it were attempted to
calculate directly $E_0^B(4q)$ from $E_0^A(4q)$ using the expression
$E_0^B(4q)=E_0^A(4q)-[E_0(B)-E_0(A)]$ \underline{after} the Bootstrap
analysis for $E_0^A(4q),E_0(A)$ and $E_0(B)$, then the separate errors on
these
three quantities would lead to a much larger uncertainty in the overall error
to be attributed to $E_0^B(4q)$.

The results are shown in table 8 using both the strategies discussed in section
2 for extracting the ground state and excited state energies $(E_{i=0,1})$.
In all cases, the three methods agree with each other -- a feature that is also
true with the other geometries. It should be added that the agreement is much
better when extracting $V_2$ or $V_4$ -- as expected, since (unlike the
$E_{i}$)
these do not involve energy differences. If a choice has to be made between
the three alternatives, then it is best to use the $E_i$ from strategy 1.
The [5,(4,3)] configuration is of particular interest, since this is simply
a square of sides $5\times 5$ tilted with respect to the basic planes.
Therefore, it can be compared with the  $5\times 5$ entry in table 7, where
it is seen that $E_{0,1}=-0.045(1), \ 0.074(4)$ -- essentially in agreement
with the corresponding values in table 8. In fact, the basic four-body
potentials $V_4(LS,5\times 5)$ and $V_4(TR, [5,(4,3)])$ are in even better
agreement having the values 1.675(2) -- see table 3 -- and 1.675(3).
The corresponding estimates from strategy 2 are 1.670(3) and 1.675(3).
Most of the difference that now appears between $E_0=-0.040(2)$ and
$E_0({\rm Wrt \ {\rm B}})=-0.046(2)$
can be attributed to the difference between $2V_2(r=5)$ and
$2V_2(x=4,y=3)$, since these are subtracted from the $V_4$'s as in
eqs.(\ref{EA},\ref{EB}).   This was seen in table 6, where
$V_2(r=5)=0.856(2)$
and $V_2(x=4,y=3)=0.859(2)$,
of which difference roughly 0.001 is expected from  a slight
lack of rotational invariance.
Another possibility -- already suggested in connection with table 4 -- is
that the
strategies used for extracting $V_{2,4}$ and, especially,  $E_i$ may be
somewhat
underestimating the errors.
\begin{table}
{Table 8:  For the TR  geometry, the values of $E_0$ and $E_1$
-- using strategy 1,  and also $E^{(0)}_M$ and $E^{(1)}_M$ -- using strategy
2 based on eq.(\ref{GT}). In addition, there is the third estimate
$E^*_0=E^{(0)}_M(4-{\rm body})-2E^{(0)}_M(2-{\rm body})$  for  $E_0$.
The errors on $E^*_i$ are simply those from $E^{(i)}_M$(4-body).
In those cases where state $B$ has a lower unperturbed energy than
state $A$,  the row "Wrt $B$" is for the binding energy calculated directly
with respect
to  state $B$ -- as in eq.(\ref{EB}).}
\begin{center}
\vskip 0.5 cm
\begin{tabular}{|c|c|c|c||c|c|c|} \hline
 $\frac{\displaystyle d}{\displaystyle a},
(\frac{\displaystyle x}{\displaystyle a},\frac{\displaystyle y}
{\displaystyle a})$ &$E_0$&$E^*_0$&$E_M^{(0)}$&$E_1$&$E^*_1$&$E_M^{(1)}$ \\
\hline
2,(1,1)&--0.1619(4)&--0.1619(8)&--0.1619(3)&0.104(1)&0.104(2)&0.104(2)\\
Wrt $B$&--0.0144(2)&--0.0148(8)&--0.0148(7)&&& \\ \hline
2,(2,1)&--0.0262(3)&--0.0270(6)&--0.0264(3)&0.1854(6)&0.182(5)&0.182(5)\\
\hline
3,(2,2)&--0.0646(6)&--0.063(1)&--0.0648(5)&0.106(1)&0.108(2)&0.106(1)\\
Wrt $B$&--0.042(1)&--0.043(6)&--0.042(1)&&& \\ \hline
3,(3,1)&--0.039(2)&--0.038(1)&--0.037(1)&0.133(1)&0.136(1)&0.133(1)\\ \hline
4,(3,2)&--0.090(1)&--0.093(9)&--0.090(1)&0.076(2)&0.078(2)&0.077(2)\\
Wrt $B$&--0.027(1)&--0.029(9)&--0.027(1)&&& \\ \hline
4,(4,1)&--0.040(1)&--0.042(3)&--0.041(1)&0.10(1)&0.097(2)&0.101(2)\\   \hline
5,(4,3)&--0.040(2)&--0.038(3)&--0.040(2)&0.074(3)&0.069(4)&0.074(3)\\
Wrt $B$&--0.046(2)&--0.042(3)&--0.046(2)&&& \\ \hline
5,(5,1)&--0.038(2)&--0.037(3)&--0.038(2)&0.082(3)&0.082(6)&0.081(3)\\  \hline
6,(5,3)&--0.051(3)&--0.046(6)&--0.051(3)&0.049(4)&0.062(4)&0.050(4)\\
Wrt $B$&--0.030(3)&--0.033(6)&--0.030(3)&&& \\ \hline
6,(6,1)&--0.029(4)&--0.026(8)&--0.029(4)&0.060(4)&0.072(6)&0.060(4)\\ \hline
\end{tabular}
\end{center}
\end{table}

\newpage

c) The Linear(L) geometry.

\vspace{0.5cm}
As demonstrated in the Appendix A, only
two of the possible basic paths $A,B,C$ in fig. 1c) can be used when extracting
the eigenvalues $E_i$ from eq.(\ref{WABC}). This is because the corresponding
3$\times$3 equation is singular due to the way the three paths are constructed.
All three combinations $A+B, \ A+C, \ B+C$ give exactly the same results
shown in the first three columns of table 9. These energies were extracted
using strategy 1 and are, in all cases, in agreement with the corresponding
 numbers from strategy 2.

 Here it is seen that the only values of $E_0$ that are
significantly different from zero come from those configurations where
the two "mesons" of length $d$ are only \underline{one} lattice spacing apart
(i.e. $r=a$) in
configuration $A$. This is unfortunate, since such configurations could
possibly suffer from lattice artefacts. However, the smallness of the other
$E_0$'s will in fact be of interest in subsequent analyses of these
numbers. On the other hand, the values of $E_1$ are quite precise, but
their interpretation is less clear than the corresponding $E_0$'s, since
they are the highest states in a variational approach and so could
possibly contain some impurities.
\begin{table}[t]
{Table 9: The energies $E_i$ for the Linear and Quadrilateral geometries as
 extracted with strategy 1. Wrt $B$ defined in table 8. Here $r=r_{23}$ in
fig. 1 and $d=r_{13}=r_{24}$.}
\begin{center}
\vskip 0.5 cm
\begin{tabular}{|c|c|c||c|c|c|} \hline
&\multicolumn{2}{c||}{ Linear}&\multicolumn{3}{c|}{Quadrilateral}\\ \hline
 $(\frac{\displaystyle d}{\displaystyle a},\frac{\displaystyle r}
{\displaystyle a})$& $E_0$& $E_1$&$E_0$& $E_1$ &$E_2$\\ \hline
(1,1)&--0.0026(1) &0.418(2)&--0.0051(3)&0.323(1)&-- \\
(1,2)&--0.0006(3)&0.735(6)&--0.0006(2)&0.65(2)&1.034(4) \\
(1,3)&0.0002(1)&1.00(1)&0.0002(2)&0.877(3)&1.03(2) \\
(2,1)&--0.0190(7)&0.202(2)&--0.0552(3)&0.112(1)&0.565(3) \\
Wrt $B$&&&--0.0451(3)&&\\ \hline
(2,2)&--0.003(1)&0.473(1)&--0.0028(3)&0.324(1)&0.733(4) \\
(2,3)&--0.0003(3)&0.688(4)&0.0008(5)&0.528(2)&0.82(1) \\
(3,1)&--0.039(1)&0.144(1)&--0.142(1)&0.072(5)&-- \\
Wrt $B$&&&--0.011(2)&&\\ \hline
(3,2)&--0.003(1)&0.377(3)&--0.01(1)&0.186(2)&0.54(1)\\
(3,3)&--0.01(1)&0.55(3)& --0.001(1)&0.36(1)&0.66(1)\\ \hline
\end{tabular}
\end{center}
\end{table}
\vskip 0.5cm

d) The Quadrilateral(Q) geometry.

\vspace{0.5cm}
The results are shown in the last three columns of table 9.
Here $d$ is the length of each meson and $r$ is the
closest distance between the two mesons.
The calculations utilized all three basic states $(A+B+C)$. However, any two
would give essentially the same results -- as explained in Appendix A. For
example, the (2,1) case gives for $E_{0,1}$ the values --0.0552(3), 0.112(1)
with $A+B$ and --0.0544(3), 0.115(1) with $B+C$.

The following comments can be made about these results:

1) The (2,1) configuration has the property that the unperturbed energies of
states $A$ and $B$ are almost equal with $E_0(A)=1.123$ and
$E_0(B)=1.114$ -- see eqs. (4.2) and (4.3). Again it
is seen that this approximate degeneracy results in a four-quark binding
energy (with respect to  $B$) of --0.0451(3), a value much larger than the
others. This is of interest, since the near degeneracy is not obvious -- unlike
the rectangle geometries (R and TR) discussed earlier. There it could be seen
directly by
arranging that the sides of the rectangles were as equal as possible in length.
However, in this case, for  state $A$ the basic flux path length is $4a$,
whereas for state $B$ it is $a+\sqrt{13}a$. Such an argument would suggest
(3,1) to be degenerate, since both paths $A$ and $B$ are then of length $6a$.
Therefore,
to get a degeneracy it is the sum of the 2-quark \underline{potentials} that
must be
equal. It is not sufficient to simply consider the linear spatial dimensions.
This would only be correct for large configurations, where all 2-quark
potentials are dominated by the linear confining potential.

2) The other state where $E_0(B)$ is seen to be less than $E_0(A)$ is the (3,1)
configuration. However, this reversal of the basic energies in fact arises for
all configurations $(d,r)$ with $d \ge 3r$.

\newpage

e) The Non-Planar(NP) geometry.

\vspace{0.5cm}

  The results are given in table 10 using the complete basis $A+B+C$.
But, as in the
Quadrilateral case -- any combination $A+B$, $A+C$ or $B+C$ gives very similar
estimates for the reasons explained in Appendix A.
\begin{table}
{Table 10:  The three energies $E_{0,1,2}$ in the Non-Planar geometry.
Wrt $B$ defined in table 8.}
\begin{center}
\vspace{0.5cm}
\begin{tabular}{|c|c|c|c|} \hline
$(\frac{\displaystyle d}{\displaystyle a},\frac{\displaystyle r}
{\displaystyle a})$
& $E_0$ & $E_1$ & $E_2$ \\ \hline \hline
(1,1) &  --0.0062(2)&0.2442(3)&0.973(4)\\
(1,2)&--&0.551(2)&1.018(4)\\
(1,3) &--0.0003(2)&0.782(7)&1.045(6) \\
(1,4) &0.0000(2)&0.997(4)&1.05(3)\\ \hline
(2,1)&--0.080(1)&0.054(1)&0.56(2)\\
Wrt $B$&--0.0178(4)&&\\ \hline
(2,2)&--0.0050(4)&0.209(3)&0.710(4)\\
(2,3)&--0.0001(3)&0.401(2)&0.749(5) \\
(2,4)&0.0010(4)&0.55(1)&0.800(5)\\        \hline
(3,1)&--0.1853(6)&0.044(1)&0.333(3)\\
Wrt $B$&--0.0048(5)&&\\ \hline
(3,2)&--0.0248(8)&0.080(1)&0.45(3)\\
(3,3)& --0.0049(7)&0.220(1)&0.59(2) \\
(3,4)& --0.001(1)&0.371(3)&0.63(1) \\          \hline
(4,1)&--0.259(1)&0.043(2)&0.16(1)\\
Wrt $B$&--0.0018(2)&&\\ \hline
(4,2)&--0.070(1)&0.034(2)&0.35(1)\\
Wrt $B$&--0.023(2)&&\\ \hline
(4,3) &--0.011(1)&0.118(2)&0.37(5)\\
(4,4)&--0.0024(3)&0.251(3)&0.50(2)\\ \hline
\end{tabular}
\end{center}
\end{table}
Here there are several cases where states $A$ and $B$ are almost degenerate
in energy, and again an increased binding energy is observed -- see
especially (2,1), (3,2) and (4,2).
\vspace{1.0cm}

\centerline{ f) Summary of results}

\vspace{0.5cm}

Fig. 3 shows a summary of how the ground state binding energy $E_0$
depends on the degree of degeneracy between states $A$ and $B$. On the
abscissa is plotted $\Delta=|E_0(A)-E_0(B)|$ and on the ordinate
axis the binding
energy with respect to the lowest non-interacting
two-body state i.e. $E_0(A)$ or $E_0(B)$.
If the results for the Tilted rectangles are normalised by the
factor $F_d=E_0(2,2)/E_0(d,d)$ to ensure each binding energy tends to
$E_0(2,2)=-0.058$ as $\Delta \rightarrow 0$, then it is found that they lie
essentially on a universal curve that drops by about an order of magnitude as
$\Delta$ changes from 0 to $\approx 0.2$. The curve also includes the (2,1)
Quadrilateral point. This rapid change was already observed in table 7, since
in going from a square($d\times d$) to a neighbouring rectangle
$\Delta =2(V_2(d+1)-V_2(d))\geq 0.2$.

\vspace{0.5cm}

\section{ Conclusion}
\setcounter{equation}{0}

In the previous sections, the binding energies of four quarks in specific
geometries have been extracted -- with the main conclusion being that the
interaction is dominated by those cases where the basic states are degenerate,
or almost degenerate, in energy. In the literature this effect has been
observed and exploited earlier. For example, in ref.\cite{Paton} it is shown
that the flux-tube model results in an interaction with a similar property.
Also, in ref.\cite{lenz} the so-called flip-flop model makes the ansatz that
the 4-quark interaction occurs only in the case of exact degeneracy. However,
the latter is developed for a purely \underline{linear} interquark potential,
which means that the interaction only occurs when -- in the notation of fig.1
-- the spatial distances $r_{13}+r_{24}$ and  $r_{14}+r_{23}$ are degenerate.
In spite of this, the present work seems -- to some extent -- to justify
this idea.

Recently -- see refs.\cite{MGP}-\cite{GBP} -- several attempts have been made
 to understand some of the above lattice results in terms of a microscopic
model based on the 2-quark potential $V_2$. In order to make these two
approaches compatible the corresponding approximations are made in both cases.
For example, in the model the quarks are static (i.e. infinite mass) and are
described in terms of mixtures of the non-orthogonal basis states
\be
A=[q_1q_3]^0[q_2q_4]^0 \ , \ B=[q_1q_4]^0[q_2q_3]^0 \ , \
C=[q_1q_2]^0[q_3q_4]^0,
\ee
where the suffix denotes a colour singlet. These states only depend on the
quark coordinates and do not contain any explicit reference to the underlying
gluon field. Because of this, the three states are linearly dependent
($C=A-B$) and satisfy the equalities $<A|B>=\frac{1}{2}=<A|C>=-<B|C>$.
It is of interest to note
that the phase appearing in the $<B|C>$ overlap must be inserted "by hand" in
eq.(\ref{WABC}), since in the lattice calculation the correlations $W^T_{ij}$
are calculated independently and, at the time,  relative phases in the
off-diagonal matrix elements ignored. This is fine for a $2\times2 $ basis,
since the
results are independent of the sign of $W^T_{12}$ -- but not so for the
$3\times 3$ case. This point is discussed in more detail in the appendix.
Now that the basis is decided, it only remains to fix the interaction
between the basis states. Being guided by success in atomic and nuclear
physics, the most obvious choice is to assume that these interactions can be
described in terms of \underline{two}--body potentials. Furthermore, if the
basis states are thought to each represent a pair of colour singlet "mesons",
then a suitable choice for the interaction is simply $\sum_{i<j}v(r_{ij})$,
where $v(r_{ij})=-\frac{1}{3}\tau_i \cdot \tau_j V_2(r_{ij})$. Here the
 $\tau$'s are the
Pauli spin matrices appropriate for SU(2) and the $V_2(r_{ij})$ are the
2-quark potentials discussed earlier and partially listed in tables 5 and 6.
With this
choice of $v(r_{ij})$, the energies of states $A$,$B$ and $C$ reduce to
$v(r_{13})+v(r_{24})$, $v(r_{14})+v(r_{23})$ and $v(r_{12})+v(r_{34})$
respectively -- as expected. The off-diagonal matrix elements have the form,
for example,
$V_{AB}=\frac{1}{2}\left(v_{13} +v_{24} +v_{14}+v_{23} - v_{12}-v_{34}
\right)$.
 Because the states $A$,$B$ and $C$ are linearly dependent,
this form of $v(r_{ij})$ results in a singular $3\times 3$ matrix, if all three
basis states are considered -- with any two giving precisely the same results.
Possibly, the reason for this is similar to the singularity occurring
in the $3\times 3$ basis in the Linear geometry -- as discussed in the
appendix.
The 4-quark energies of this simple model can then be obtained by diagonalizing
\be
\label{Ham}
\left[{\bf V}-\lambda_i {\bf N}\right]\Psi_i=0,
\ee
with
\be
\label{NV}
{\bf N}=\left(\begin{array}{ll}
1&1/2\\
1/2&1\end{array}\right)\ \ {\rm and}\ \ {\bf V}=\left(\begin{array}{cc}
v_{13}+v_{24} & V_{AB}\\
V_{BA}&v_{14}+v_{23}\end{array}\right),
\ee

In analogy with eqs.(\ref{diff}) and (\ref{Ei}), the eigenvalues are expressed
 with
respect to state $A$, since this usually has the lowest energy, i.e.
\be
E'_i=\lambda_i-E_0(A).
\ee
These $E'_i$ should now represent a model for the lattice results $E_i$.
For the Rectangular and Large Square geometries, the $E'_{0,1}$ are given in
table 7 using in eq.(\ref{Ham}) the global fit for the $V_2(r_{ij})$ from
table 5.
There it is seen that this is a \underline{poor} model -- with $E'_0$ being
much more attractive than $E_0$ and $E'_1$ much more repulsive than $E_1$. This
is a general feature seen in all other geometries.
In refs.\cite{MGP}-\cite{GBP},  this simple model is extended by introducing a
phenomenological function $f$ that is interpreted as a gluon field overlap
factor such that $<A|B>\rightarrow\frac{f}{2}$ and $V_{AB}\rightarrow fV_{AB}$.
   The reader should
refer to these references for more details, since the main purpose of this
paper is to present the \underline{model-independent} results from the Monte
Carlo lattice simulations. The only reason for discussing the failure of the
simple model in eq.(\ref{Ham}) is because it is essentially the basis of many
attempts to calculate the interaction between multi-quark clusters
e.g. refs.\cite{MMS} and \cite{NN}. These all assume that the interaction can
be
    reduced to
a sum of purely 2-quark potentials. This sum is then introduced into some
many-body formalism such as the Resonating Group technique -- that has proved
successful for calculating the potential between atoms and between nuclei --
to give a non-local meson-meson or nucleon-nucleon potential. It should be
added that the present criticism is not against these many-body techniques,
since they can be extended to include the effect of the multi-quark factor $f$
-- see refs.\cite{Masud1} and \cite{Masud2}. However, there seems to be no
justification
for assuming that the input to these many-body techniques can be expressed
in terms of purely 2-quark interactions.

It should be emphasised that the 4-quark energies tabulated in section 4 are
-- for the static quenched limit in SU(2) -- \underline{model-independent} and
\underline{exact}, within the accuracy of the Monte Carlo
simulation. This being the case, other methods for calculating these energies
e.g. quark clusters \cite{MMS}-\cite{NN} and light-cone models \cite{light}
should result in the \underline{same}  values. Any failure to do so should be
interpreted as a weakness of those models.

The authors wish to thank Dr. J.E. Paton for many useful comments at all stages
of this work. They also  wish to acknowledge that these  calculations were
performed on the CRAY X-MP machine at  Helsinki.

\appendix

\chapter{Bases for 4-quark systems}

\renewcommand{\theequation}{A.\arabic{equation}}
\vspace{0.5cm}


Combining together multi-quark states into colour singlet combinations is
straightforward in principle but care must be taken with several small details.
To illustrate the analysis  needed for 4-quark systems, first consider the
simpler case of two quarks. With SU(2) colour, the two-quark system will be a
baryon. However, the baryon has the same features as the quark-antiquark meson
state. This arises because the quark and antiquark both transform under the
same 2-dimensional representation of the colour group. To expose this,
 the colour charge conjugation operator will be needed and it is taken as
$\epsilon_{ij}$, with the definition $\epsilon_{12}=-\epsilon_{21}=1,
\epsilon_{11}=\epsilon_{22}=0$. Here  the quarks are static so they only have a
colour index and can be represented as a quark at $x$ as $q_i(x)$.  Then  the
antiquark
field is given  in terms of the quark field by

\begin{equation}
\overline{q_i(x)}=-\epsilon_{ij} q_j(x).
\end{equation}

\noindent
To make gauge invariant operators from quarks at different spatial positions,
it is necessary to include a path-ordered integral to relate the colour
coordinate systems at the two spatial points. In lattice gauge theory these
path ordered integrals are provided by the ordered product of links  and in
this
case they are SU(2) matrices $P_{ij}(z,x)$. They are to be thought  of as a
colour flux between two points and are generally referred to as paths. The
choice of spatial position $z$ at which to make this colour reference is
arbitrary. Then the two-quark operator is

\begin{equation}
T_{ij} = P_{ir}(z,x) q_r(x) P_{js}(z,y) q_s(y) |0>.
\end{equation}

\noindent
Now only colour singlet combinations of two quarks will have non-zero vacuum
expectation values. There is only one such combination and it is given by
contracting with $\epsilon_{ij}$. Inserting the identity
$-\epsilon_{ut} \epsilon_{tr}=\delta_{ur}$ gives

\begin{equation}
S=\epsilon_{ij} T_{ij} = - q_u(x) \epsilon_{ut} \epsilon_{tr}
 P_{ir}(z,x) \epsilon_{ij}  P_{js}(z,y) q_s(y) |0>.
\end{equation}

\noindent
Then using the identity for SU(2) matrices that
$\epsilon_{ij} P_{jk} \epsilon_{kl}=-P^{*}_{il}$ gives

\begin{equation}
S = q_u(x) \epsilon_{ut} P^{\dag}_{tj}(z,x) P_{js}(z,y) q_s(y) |0>
 = \overline{q_t(x)} P_{ts}(x,y;z) q_s(y) |0>.
\end{equation}

\noindent
Here $P(x,y;z)$ is the colour flux path from $y$ to $x$ via $z$.  To obtain the
energy  of this two-quark state then needs an evaluation in the lattice vacuum
samples of $<0|S^{\dag}(0) S(t)|0>$. This clearly corresponds to evaluating a
closed flux loop (a Wilson loop) joining the 4 points $ (y,0), (x,0), (x,t),$
and $(y,t)$.
 This result is usually assumed without the detailed discussion given
above, but some of the discussion will be less trivial for 4-quark systems.
Note that there is considerable freedom to choose the path from $x$ to $y$
at both time 0 and $t$.
 This freedom can be used to construct a variational
approach using several such paths.

Returning now to the 4-quark system, the same approach leads to a 4-quark
operator given by

\begin{equation}
T_{ijkl} = P_{ir}(z,x_1) q_r(x_1) P_{js}(z,x_2) q_s(x_2)
           P_{kt}(z,x_3) q_t(x_3) P_{lu}(z,x_4) q_u(x_4) |0>.
\end{equation}

\noindent
Now colour singlet states can be obtained by pairing the quarks. There are
three ways to achieve this corresponding to contracting $T$ with tensors

\begin{equation}
A_{ijkl}=\epsilon_{ij} \epsilon_{kl}, \
B_{ijkl}=\epsilon_{il} \epsilon_{kj} \ \hbox{ and} \
C_{ijkl}=\epsilon_{ik} \epsilon_{jl}
\end{equation}

\noindent
These three tensors correspond to the pairings $A$, $B$ and $C$ introduced
in fig.1.

As in the two-quark case above, the energy eigenstates can be obtained by
evaluating the correlation of this 4-quark operator at times 0 and $t$. Again
these correlations can be cast into Wilson loops. There are now  three
possible tensorial structures at each time value. The diagonal correlations
($AA$, $BB$ or $CC$) will each
yield two Wilson loops, while the off-diagonal correlations
will yield one Wilson loop. This latter  Wilson loop has to be multiplied by
non-trivial phase factors that arise because of the antisymmetry of the
$\epsilon$-factors. Indeed, with the present convention, the correlation of
tensors $B$ and $C$ has an extra minus sign.

These three tensors $A$, $B$ and $C$ are linearly dependent with $A=B+C$ as can
be seen by using the identity that

\begin{equation}
\epsilon_{ij} \epsilon_{kl} = \delta_{ik} \delta_{jl}-\delta_{il} \delta_{jk}.
\end{equation}

\noindent
Another way to appreciate this is that there are only two independent ways to
couple 4 spin-half particles to a scalar.

This linear dependence of the tensors will only imply that there is a linear
dependence of paths $A$, $B$ and $C$ if the latter are indeed constructed
exactly in
the way defined here. Namely  a common  point $z$ must exist such that  paths
to each of the 4 quarks from $z$ are used to create path combinations $A$, $B$
and $C$. In the present work, this is exactly the case for the linear
geometry as can be seen from the detailed construction of the paths.

In other cases, the path definitions do not correspond exactly and one can use
all three possible tensor structures to make 4-quark states. These can then be
employed as a variational basis. In practice, it turns out that in some cases
such as  the quadrilateral geometry, there is  an approximate degeneracy of the
3 path combinations $A$, $B$ and $C$ so little is gained by including the
third path.

Since  there are two independent tensorial combinations of 4 quarks, it might
be thought  useful to separate the observed energy eigenstates according to
which is which. For instance to explore whether quarks 1 and 2 are combined in
a scalar ($A$) or a vector ($B-C$). Such a separation depends on the paths
used to link each quark to the reference point $z$ and so is not of physical
significance except in the weak coupling limit where all gluon fields are small
and all path matrices are the identity.

\newpage
{\bf Figure Captions}

\vskip 0.5cm

\noindent Fig.1 The basic four-quark geometries

\hspace*{1.0cm} a) Rectangle(R) and Large Squares(LS)

\hspace*{1.0cm} b) Tilted Rectangle(TR)

\hspace*{1.0cm} c) Linear(L)

\hspace*{1.0cm} d) Quadrilateral(Q)

\hspace*{1.0cm} e) Non-Planar(NP)

\vskip 0.5 cm

\noindent Fig.2 The two paths $P_i(a)$ and $P_i(b)$, of fuzzing level $i$,
used in the
construction of the L-shaped two-quark potential $V_2(x,y)$
\vskip 0.5 cm

\noindent Fig.3  The binding energies $E_0$ -- with respect to the
lower of $E_0(A)$ and $E_0(B)$ -- as a function of  the difference
$\Delta=|E_0(A)-E_0(B)|$ between the basis
energies $E_0(A)$ and $E_0(B)$. All energies are in units of $1/a$.

Solid line: Tilted Rectangle  case

Dashed line: Non-Planar  case

Crosses: Quadrilateral case


\begin{thebibliography}{99}
\bibitem{MMS}
T. Barnes and E.S. Swanson, Phys. Rev. D46 (1992) 131;

D. Blaschke and G. R\"{o}pke, Phys. Lett. B299 (1993) 332.
\bibitem{NN}
M. Oka and K. Yazaki, International Review of Nuclear Physics -- Vol. 1,
Quarks and Nuclei (World Scientific, Singapore, 1984) p. 489;

Y. Yamauchi, K. Tsushima and A. Faessler,  Few-Body Systems 12 (1992) 69 --
for a list of references on this topic;

T. Barnes, S. Capstick, M.D. Kovarik and E.S. Swanson, Phys. Rev. C48
(1993) 539.
\bibitem{MGP}
O. Morimatsu, A.M. Green and J. Paton, Phys. Lett. B258 (1991) 257.
\bibitem{GMP}
A.M. Green, C. Michael and J. Paton, Phys. Lett. B280 (1992) 11.
\bibitem{GMP1}
A.M. Green, C. Michael and J.E. Paton, Nucl. Phys. A554 (1993) 701.
\bibitem{GMP2}
\mbox{A.M. Green, C. Michael, J.E. Paton and M.E. Sainio},
Int. J. Mod. Phys. E2 (1993) 479.
\bibitem{GBP}
S. Furui, A.M. Green, B. Masud and J.E. Paton, in preparation.
\bibitem{Markum}
K. Rabitsch, H. Markum and W. Sakuler, Phys. Lett. B318 (1993) 507.
\bibitem{Vinhmau}
R. Vinh Mau, C. Semay, B. Loiseau and M. Lacombe, Phys. Rev. Lett. 67
(1991) 1392.
\bibitem{CM1}
S. Perantonis, A. Huntley and C. Michael, Nucl. Phys. B326 (1989) 544.
\bibitem{CM2}
S. Perantonis and C. Michael, Nucl. Phys. B347 (1990) 854.
\bibitem{Hunt}
A. Huntley and C. Michael, Nucl. Phys. B270 (1986) 123.
\bibitem{chrism}
C. Michael, Nucl. Phys. B (Proc. Suppl.) 17 (1990) 59;

C. Michael, Phys. Lett. B232 (1989) 247.
\bibitem{Boot}
C.A. Whitney, Random Processes in Physical Systems (John Wiley, New York, 1990)
p. 254.
\bibitem{cdata}
B.A. Berg, Comput. Phys. Comm. 69 (1992) 7;

C. Michael, Phys. Rev. D49 (1994) 2616.
\bibitem{LL}
L. Lyons, Statistics for Nuclear and Particle Physicists  (Cambridge University
Press, Cambridge, 1986) p. 13.
\bibitem{Roos}
F. James and M. Roos, Comput. Phys. Comm. 10 (1975) 343.
\bibitem{Paton}
A.M. Green and J. Paton, Nucl.Phys. A492 (1989) 595.
\bibitem{lenz}
F. Lenz et al. , Ann. Phys. (N.Y.) 170 (1986) 65;

K. Masutani, Nucl.Phys. A468 (1987) 593.
\bibitem{Masud1}
B. Masud, J. Paton, A.M. Green and G.Q. Liu, Nucl. Phys. A528 (1991) 477.
\bibitem{Masud2}
B. Masud, A Model for $q^2\bar{q}^2$ Systems, Illustrated by an
Application to
$K\bar{K}$ Scattering , Helsinki University preprint HU-TFT-93-30, appears
in Hep-ph/9401336.
\bibitem{light}
S.J. Brodsky, G. McCartor, H.C. Pauli and S.S. Pinsky,
Particle World 3 (1993) 109.
\vskip 0.5cm

\end{thebibliography}
\end{document}